\documentclass[twocolumn]{jpsj2} %% two-column layout
\usepackage{bm}

\title{ Doping Dependence of Normal-State Properties in Iron-Based Oxypnictide Superconductor LaFeAsO$_{1-y}$ Probed by $^{57}$Fe-NMR and $^{75}$As-NMR/NQR}

\author{
Hidekazu Mukuda$^{1,4}$\thanks{E-mail address: mukuda@mp.es.osaka-u.ac.jp},  Nobuyuki Terasaki$^{1}$, Nobukatsu Tamura$^{1}$, Hiroaki Kinouchi$^{1}$, Mitsuharu Yashima$^{1,4}$, Yoshio Kitaoka$^{1}$, Kiichi Miyazawa$^{2}$, Parasharam M. Shirage$^{2}$, Shinnosuke Suzuki$^{3}$, Shigeki Miyasaka$^{3,4}$,\\ Setsuko Tajima$^{3,4}$, Hijiri Kito$^{2,4}$, Hiroshi Eisaki$^{2,4}$, and Akira Iyo$^{2,4}$ 
}

\inst{
$^{1}$Graduate School of Engineering Science, Osaka University, Toyonaka, Osaka 560-8531 \\
$^{2}$National Institute of Advanced Industrial Science and Technology (AIST), Umezono, Tsukuba 305-8568 \\
$^{3}$Department of Physics, Graduate School of Science, Osaka University, Toyonaka, Osaka 560-0043 \\
$^{4}$JST, TRIP (Transformative Research-Project on Iron Pnictides), Chiyoda, Tokyo 102-0075 
}

\abst{
We report systematic $^{57}$Fe-NMR and $^{75}$As-NMR/NQR studies on an underdoped sample ($T_c=20$ K), an optimally doped sample ($T_c=28$ K), and an overdoped sample ($T_c=22$ K) of oxygen-deficient iron (Fe)-based oxypnictide superconductor LaFeAsO$_{1-y}$.  A microscopic phase separation between superconducting  domains and magnetic domains is shown to take place in the underdoped sample, indicating a local inhomogeneity in association with the density distribution of oxygen deficiencies. 
As a result, $1/T_1T$ in the normal state of the superconducting domain decreases significantly upon cooling at both the Fe and As sites regardless of the electron-doping level in LaFeAsO$_{1-y}$. 
On the basis of this result, we claim that $1/T_1T$ is not always enhanced by antiferromagnetic fluctuations close to an antiferromagnetic phase in the underdoped superconducting sample. This contrasts with the behavior in hole-doped Ba$_{0.6}$K$_{0.4}$Fe$_2$As$_2$($T_c= 38$ K), which exhibits a significant increase in $1/T_1T$ upon cooling.  We remark that the crucial difference between the normal-state properties of LaFeAsO$_{1-y}$ and Ba$_{0.6}$K$_{0.4}$Fe$_2$As$_2$ originates from the fact that the relevant Fermi surface topologies are differently modified depending on whether electrons or holes are doped into the FeAs layers.
} 

\kword{superconductivity, iron-based oxypnictide, LaFeAsO, NMR}

\begin{document}

\maketitle

\date{\today}

%%%%%%%%%%%%%%%%%%%     Introduction     %%%%%%%%%%%%%%%%%%%%%%%%%%%%%%%%%%

\section{Introduction}

\begin{figure}[tbp]
\begin{center}
\includegraphics[width=0.8\linewidth]{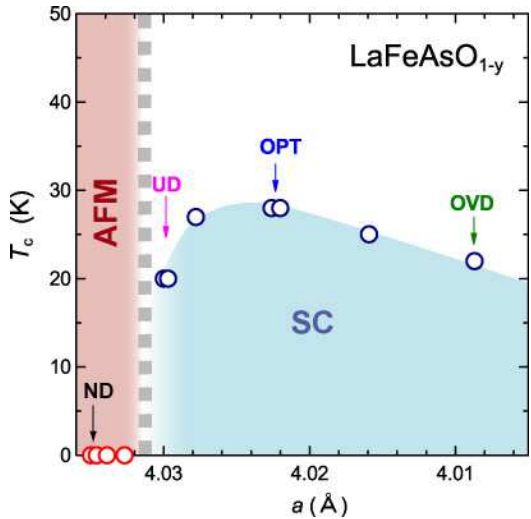}
\end{center}
\caption[]{(color online)
Phase diagram of LaFeAsO$_{1-y}$ as a function of $a$-axis length evaluated at room temperature. An AFM-to-SC transition appears to take place at an $a$-axis length of approximately $a\simeq 4.03 \dot{A}$, since bulk SC was not obtained in the sample with $a> 4.03 \dot{A}$. In the SC phase,  note that even though the nominal oxygen content is different, $T_c$ is almost the same if the $a$-axis length is identical. 
}
\label{phasediagram}
\end{figure}

The discovery of superconductivity (SC) in the iron (Fe)-based oxypnictide  LaFeAsO$_{1-x}$F$_x$ with   SC transition temperature $T_c$=26 K has attracted considerable interest in the fields of condensed-matter physics and materials science \cite{Kamihara2008}. Soon after this discovery, the $T_c$ of LaFeAsO$_{0.89}$F$_{0.11}$ was reported to increase up 43 K upon applying pressure \cite{Takahashi}, and the replacement of the La site by other rare-earth (RE) elements significantly increases $T_c$ to more than 50 K \cite{Ren1,Kito,Ren2,GdFeAsO}. The structure of the mother material contains alternately  stacked RE$_2$O$_2$ and Fe$_2$As$_2$ layers along the $c$-axis, where the Fe atoms of the FeAs layer are located in a fourfold coordination forming a FeAs$_{4}$ tetrahedron. The mother material LaFeAsO is a semimetal with a stripe antiferromagnetic (AFM) order with ${\bf Q}=(0,\pi)$ or $(\pi,0)$ \cite{Cruz}. The substitution of fluorine for oxygen and/or oxygen deficiencies in the LaO layer yields a novel SC \cite{Kamihara2008,Ren1,Kito,Ren2,GdFeAsO}.  Remarkably, Lee {\it et al.} found that $T_{c}$ increases to a maximum value of 54 K when the FeAs$_{4}$ tetrahedron is transformed into a regular tetrahedron\cite{C.H.Lee}.  
Related to this fact, Shirage {\it et al.} revealed that the $a$-axis length, corresponding to the distance among Fe atoms in the square lattice, also has a strong correlation with $T_c$ in the RE-Fe-As-O system, as presented in Fig.~\ref{phasediagram}\cite{Shirage}. 

In the normal state of LaFeAsO$_{1-x}$F$_x$, pseudogap-like behavior was first reported on the basis of photoemission \cite{Sato}, $^{75}$As-NMR,\cite{Nakai} and $^{19}$F-NMR\cite{Ahilan} measurements. Similar behavior was also reported by $^{75}$As-NMR probe for the electron-doped superconductor Ba(Fe$_{0.895}$Co$_{0.105}$)$_{2}$As$_{2}$\cite{Ning}, but not for the hole-doped Ba$_{1-x}$K$_{x}$Fe$_{2}$As$_{2}$ \cite{FukazawaSC,MukudaPhysC}. This suggests that the decrease of $^{75}$As-NMR-$(1/T_1T)$ upon cooling, or pseudogap-like behavior, may characterize the normal-state properties in electron-doped Fe-pnictide systems. However, it has been reported by Nakai {\it et al.} \cite{Nakai,Nakai2} that $^{75}$As-NMR-($1/T_1T$) is markedly enhanced upon cooling for underdoped LaFeAsO$_{1-x}$F$_{x}$ ($x=0.04$). In addition, they observed a peak at $\sim$30 K in $1/T_1$, below which the $^{75}$As-NMR spectral width  starts to broaden gradually upon cooling.  

In this paper, we report systematic studies on the doping dependence of the normal-state properties of electron-doped superconductors LaFeAsO$_{1-y}$ by means of $^{57}$Fe-NMR and $^{75}$As-NMR/NQR measurements. A microscopic phase separation between SC and magnetic domains is demonstrated to take place in an underdoped sample, indicating a local inhomogeneity in association with the density distribution of oxygen deficiencies. As a result, it is highlighted that $1/T_1T$ at both the Fe and As sites decreases markedly upon cooling, even in SC domains close to an AFM phase, and hence this behavior is a common feature regardless of electron-doping level. This, however, differs from the drastic doping dependence previously observed for $^{75}$As-NMR-$(1/T_1T)$, which is markedly enhanced upon cooling in the underdoped sample \cite{Nakai2}. 
We also address the crucial difference observed between the normal-state properties of LaFeAsO$_{1-y}$ and Ba$_{0.6}$K$_{0.4}$Fe$_2$As$_2$. 

%----------------------- Fig.1 phase diagram ---------------------
%----------------------------------------------------------------------
%---------------------------------------------------------------
\begin{table}[htbp]
\caption{$T_c$ and lattice parameters at room temperature for a nondoped sample (ND), an underdoped sample (UD), an optimally doped sample (OPT), and an overdoped sample (OVD) of LaFeAsO$_{1-y}$. Note that $T_c$ can be almost exactly reproduced when the $a$-axis length is identical, even though the nominal oxygen content is different. 
}
\begin{center}
\begin{tabular}{ccccccc}
\multicolumn{2}{c}{Sample} &  $T_{c}$ [K] & $a$ [\AA ] & c [\AA ] & References\\ \hline
ND & LaFeAsO$^\dagger$        & -$^*$ & 4.0346 & 8.752 &  \\ \hline
UD & LaFeAsO$_{0.78}^\dagger$ & 20 & 4.0300 & 8.7136 &  \\ 
   & LaFeAsO$_{0.75}$         & 20 & 4.0297 & 8.7143 & \\ \hline
OPT & LaFeAsO$_{0.7}^\dagger$ & 28 & 4.0226 & 8.7065 & \cite{Terasaki}\\ 
    & LaFeAsO$_{0.6}$         & 28 & 4.0220 & 8.7110 & \cite{Mukuda}\\ \hline
OVD & LaFeAsO$_{0.6}^\dagger$ & 22 & 4.0087 & 8.6978 & \\ \hline
\end{tabular}
\end{center}
\footnotesize{$\dagger$)$^{57}$Fe-enriched sample.}\\ 
\footnotesize{$*$)$T_{\rm N}\sim$ 145 K.}\\ 
\label{tb:sample}
\end{table}
%-------------------------------------------------------------------------

%%%%%%%%%%%%%%%%%%%%    Experimental    %%%%%%%%%%%%%%%%%%%%%%%%%%%%%
\section{Experimental}

Polycrystalline samples of LaFeAsO$_{1-y}$ were synthesized via a high-pressure synthesis technique described elsewhere \cite{Kito}. In particular, starting materials of LaAs, $^{57}$Fe, and $^{57}$Fe$_{2}$O$_{3}$ enriched by the $^{57}$Fe isotope ($^{57}$Fe: nuclear spin $I$ = 1/2, $^{57}\gamma_n/2\pi= 1.3757$ MHz/T) were mixed at a nominal composition of LaFeAsO$_{1-y}$. Powder X-ray diffraction measurements indicate that these samples are almost entirely composed of a single phase, even though the oxygen content of the samples differs from the nominal content, depending on the oxidation of the starting RE elements. In fact, the SC transition temperature ($T_c$) for all samples was uniquely determined by a susceptibility measurement, which exhibits a marked decrease due to the onset of SC diamagnetism. There is a strong correlation between $T_c$ and the $a$-axis length in RE-Fe-As-O systems\cite{Shirage}.  As summarized in Table~I and Fig.~\ref{phasediagram}, the $T_c$ of LaFeAsO$_{1-y}$ is almost the same if the $a$-axis length is identical regardless of the process used to prepare samples. Bulk SC was not obtained in the sample with $a> 4.03 \dot{A}$. $^{57}$Fe-NMR  and $^{75}$As-NMR/NQR measurements were carried out on an underdoped sample (UD), an optimally doped sample (OPT), and an overdoped sample (OVD) with $T_c=$ 20 K, 28 K, and 22 K, respectively. All samples were moderately crushed into powder.  The nuclear spin-lattice relaxation rate $1/T_1$ was measured by the saturation-recovery method.

\section{Results}

\subsection{$^{57}$Fe-NMR study on LaFeAsO$_{1-y}$}

\begin{figure}[tbp]
\begin{center}
\includegraphics[width=0.8\linewidth]{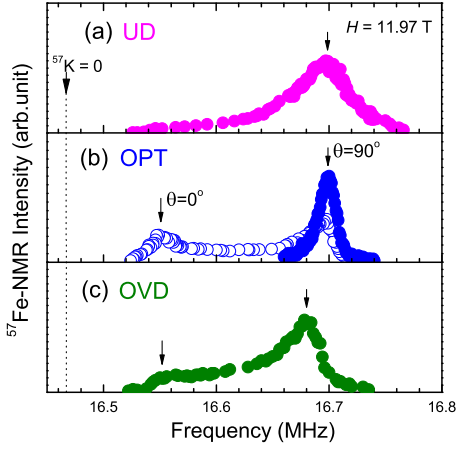}
\end{center}
\caption[]{(color online)
$^{57}$Fe-NMR spectra for (a) UD, (b) OPT, and (c) OVD  of LaFeAsO$_{1-y}$ at $H=$ 11.97 T and 30 K. For OPT, which was oriented along the direction including the $ab$-plane, the spectra with a single narrow peak and two horn-shaped peaks respectively correspond to fields parallel and perpendicular to the direction including the $ab$-plane. The broad spectrum of UD (a) with the field parallel to the $ab$-plane originates from the inhomogeneous distribution of the $^{57}$Fe Knight shift, which is caused by the phase separation into SC and magnetic domains at a microscopic scale in UD close to an AFM phase (see text). The spectrum of OVD (c), which was not oriented, indicates an asymmetric powder pattern due to the anisotropy of the $^{57}$Fe Knight shift. }
\label{fig:FeNMRspectrum}
\end{figure}

\begin{figure}[htbp]
\begin{center}
\includegraphics[width=0.9\linewidth]{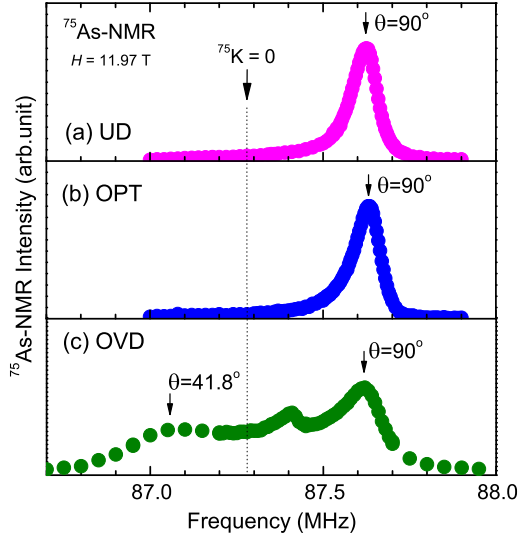}
\end{center}
\caption[]{(color online)
$^{75}$As-NMR spectra for (a) UD, (b) OPT, and (c) OVD of LaFeAsO$_{1-y}$ at $H$=11.97 T and 30 K. The sharp peak at approximately $f\sim$ 87.6 MHz originates from the central transition ($I=+1/2\leftrightarrow -1/2$) in the field parallel to the $ab$-plane ($\theta=90^\circ$). The asymmetric lineshape in the OVD  sample is attributed to the unoriented powder sample. It is noteworthy that this resonance frequency for $\theta=90^\circ$ does not vary for all the samples owing to the much smaller Knight shift at the As site than at the Fe site.
}
\label{fig:AsNMRspectra}
\end{figure}

Figure~\ref{fig:FeNMRspectrum} shows $^{57}$Fe-NMR spectra obtained by sweeping the frequency ($f$) at magnetic field $H=$ 11.97 T at 30 K for (a) UD, (b) OPT\cite{Terasaki}, and (c) OVD of LaFeAsO$_{1-y}$.  
For OPT, which was oriented along the direction including the $ab$-plane, we observed a single narrow peak when $H$ was applied parallel to the orientation direction, as shown in Fig.~\ref{fig:FeNMRspectrum}(b)\cite{Terasaki}. In the field perpendicular to the orientation direction, two horn-shaped peaks were observed, which arise from crystals with $\theta=90$ and $0^\circ$, where $\theta$ is the angle between the field and the $c$-axis of the crystal. Such a spectral shape originates from the anisotropy of the Knight shift with $^{57}K^\perp \sim 1.413$\% and $^{57}K^\parallel \sim 0.50$\% at 30 K for $\theta=90$ and $0^\circ$, respectively\cite{Terasaki}.  For UD, the $^{57}$Fe-NMR spectral width at 30 K is broader than that for OPT, even though it is as narrow as that for OPT at 200 K (see Fig. \ref{fig:linewidth}(a)). This suggests that the broad spectrum at 30 K for UD with the field parallel to the $ab$-plane is ascribed to the distribution of the $^{57}$Fe-NMR  Knight shift. Figure \ref{fig:FeNMRspectrum}(c) shows the spectrum of nonoriented OVD, which has an asymmetric powder pattern due to the anisotropy of the $^{57}$Fe-NMR Knight shift, as  corroborated by the $^{75}$As-NMR spectrum (see Fig.~\ref{fig:AsNMRspectra}(c)). 
The temperature ($T$) dependence of the $^{57}$Fe Knight shift was precisely measured for OPT as reported in the literature \cite{Terasaki}, but not for UD and OVD because of their broad NMR spectral width. Thus, we focus on the nuclear relaxation behavior to unravel the doping dependence of the normal-state properties in LaFeAsO$_{1-y}$.

%------------------- Fig.2 Fe NMR spectrum ---------------------
%----------------------------------------------------------------------
%--------------------  Fig.3 Recovery curve  ---------------------------
\begin{figure}[htbp]
\begin{center}
\includegraphics[width=0.8\linewidth]{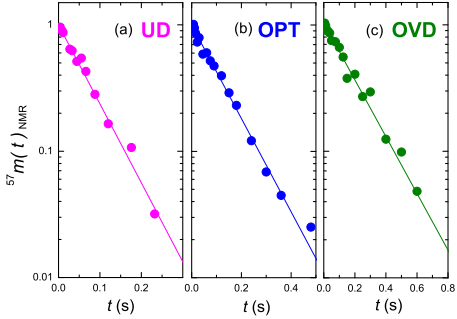}
\end{center}
\caption[]{(color online)
Recovery curves of $^{57}$Fe nuclear magnetization in the normal state at $T$= 80 K and $H$= 11.97 T for (a) UD, (b) OPT, and (c) OVD of LaFeAsO$_{1-y}$. In the entire range of $T$, the recovery curves are almost uniquely fitted by $^{57}m(t)_{\rm NMR}$, as indicated by solid lines.
}
\label{Ferecovery}
\end{figure}
%----------------------------------------------------------------------

The $^{57}(1/T_1)$ of $^{57}$Fe was measured at $H$=11.97 T along the $ab$-plane ($\theta=90^\circ$) for all samples.  As shown in Fig.~\ref{Ferecovery}, $^{57}T_1$ was almost uniquely determined over the entire $T$ range from a single exponential function of $^{57}m(t)_{\rm NMR}$, which is given by 
\[
^{57}m(t)_{\rm NMR}\equiv\frac{M(\infty)-M(t)}{M(\infty)}=\exp\left(-\frac{t}{T_{1}}\right),
\]
where $M(\infty)$ and $M(t)$ are the respective nuclear magnetizations for the thermal equilibrium condition and at time $t$ after the saturation pulse.

%**************  Fig. 4****************************************
\begin{figure}[htbp]
\begin{center}
\includegraphics[width=0.8\linewidth]{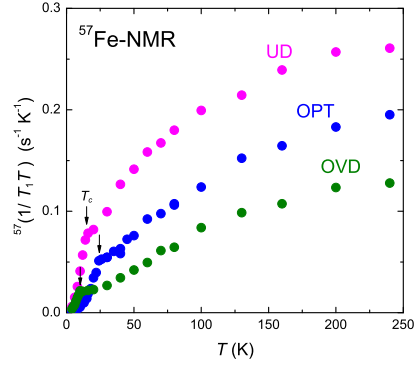}
\end{center}
\caption[]{(color online)
$T$ dependence of $^{57}$Fe-$(1/T_{1}T)$ for LaFeAsO$_{1-y}$. In the normal state, $1/T_1T$ decreases significantly upon cooling for all samples, and the magnitude of $1/T_1T$ decreases markedly as the doping level increases when going from UD to OVD. The relaxation behaviors of $^{57}$Fe-$(1/T_{1}T)$ for OPT and OVD respectively  resemble those of $^{75}$As-NMR-$(1/T_{1}T)$ for LaFeAsO$_{1-x}$F$_{x}$ with $x=0.11$ and 0.14 \cite{Nakai,Nakai2}. 
In contrast, the relaxation behavior of $^{57}$Fe-$(1/T_{1}T)$ for UD is considerably different from that of $^{75}$As-NMR-$(1/T_{1}T)$ for the underdoped LaFeAsO$_{1-x}$F$_{x}$ with $x=$0.04, which is enhanced upon cooling \cite{Nakai,Nakai2}. 
}
\label{fig:FeNMRT1}
\end{figure}
%****************************************************************************

Figure~\ref{fig:FeNMRT1} shows the $T$ dependences of $^{57}$Fe-$(1/T_1T)$ for UD, OPT,\cite{Terasaki} and OVD.  In the normal state, $1/T_1T$ decreases significantly upon cooling for all samples, and the magnitude of $1/T_1T$ decreases markedly as the doping level increases from UD to OVD. 
The relaxation behaviors of $^{57}$Fe-$(1/T_{1}T)$ for OPT and OVD respectively resemble those of $^{75}$As-NMR$(1/T_{1}T)$ for LaFeAsO$_{1-x}$F$_{x}$ with $x=0.11$ and 0.14 \cite{Nakai,Nakai2}. 
In contrast, the relaxation behavior of $^{57}$Fe-$(1/T_{1}T)$ for UD is considerably different from that of $^{75}$As-NMR-$(1/T_{1}T)$ for the underdoped LaFeAsO$_{1-x}$F$_{x}$ with $x=$0.04, which is enhanced upon cooling \cite{Nakai,Nakai2}. 
In order to reconcile why the $T$ dependence of $1/T_1T$ for the present $^{57}$Fe-NMR differs from the previous results obtained by $^{75}$As-NMR measurements, in the next section we discuss with the $^{75}$As-NMR $T_1$ measurements on the same samples as those in $^{57}$Fe-NMR study. 

%**************  Fig. 5**************************************************
%****************************************************************************

\subsection{$^{75}$As-NMR/NQR studies on LaFeAsO$_{1-y}$}

Figures~\ref{fig:AsNMRspectra}(a)-\ref{fig:AsNMRspectra}(c) respectively show the $^{75}$As-NMR spectra at 30 K and $H=$ 11.97 T for UD, OPT, and OVD of LaFeAsO$_{1-y}$.  The sharp peak at approximately $f\sim$ 87.6 MHz for all samples originates from the central transition ($I=+1/2\leftrightarrow -1/2$) in the field parallel to the $ab$-plane ($\theta=90^\circ$). 
The spectrum of OVD (c), which was not oriented, indicates an asymmetric powder pattern due to the anisotropy of the $^{75}$As Knight shift. Note that the resonance frequency for $\theta=90^\circ$ is almost invariant for all samples, in contrast with that in the $^{57}$Fe-NMR spectra indicated in Fig.~\ref{fig:FeNMRspectrum}.
This is because that the $^{75}$As Knight shift is smaller than the $^{57}$Fe Knight shift.

\begin{figure}[htbp]
\begin{center}
\includegraphics[width=0.9\linewidth]{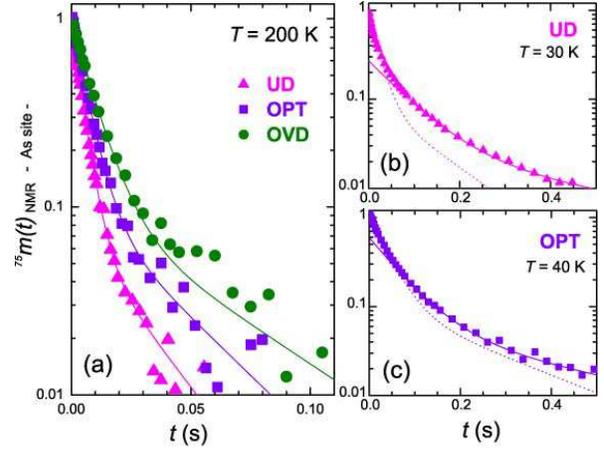}
\end{center}
\caption[]{(color online)
Recovery curves of $^{75}$As nuclear magnetization. (a) The curves are uniquely determined by the theoretical function $^{75}m(t)_{\rm NMR}$ (solid lines) above $\sim$160 K for all samples. However, below $\sim$150$\pm$10 K, $^{75}m(t)_{\rm NMR}$ cannot be reproduced by the theoretical curve in (b) UD and (c) OPT because of a phase separation (see text).  The volume fraction with a short $T_1$ is increased when a sample approaches an AFM order and/or a structural transition. 
}
\label{fig:Asrecovery}
\end{figure}

The recovery curve of $^{75}$As nuclear magnetization ($I=3/2$) obtained by the $^{75}$As-NMR measurement is expressed by the following theoretical curve: 
\[
^{75}m(t)_{\rm NMR}=0.1\exp\left(-\frac{t}{T_{1}}\right) + 0.9\exp\left(-\frac{6t}{T_{1}}\right).
\]
As shown in Fig.~\ref{fig:Asrecovery}(a), $^{75}m(t)_{\rm NMR}$ can be fitted by a theoretical curve at high temperatures ($T\ge 160$ K) for all samples. 
For UD, however, since it cannot be reproduced by a single component of $T_1$ below $\sim$150$\pm$10 K, as shown in Fig.~\ref{fig:Asrecovery}(b), a long component ($T_{1L}$) and a short component ($T_{1S}$) are tentatively estimated by assuming the expression  $^{75}m(t)_{\rm NMR}\equiv A_S^{75}m_{S}(t)_{\rm NMR}+A_L^{75}m_{L}(t)_{\rm NMR}$. Here, $A_S$ and $A_L$ (with $A_S+A_L=1$) are the respective fractions of domains with $T_{1S}$ and $T_{1L}$. The respective $A_S$ for UD, OPT, and OVD are 0.7$-$0.8, 0.3$-$0.4, and $\sim$0, revealing that the volume fraction of domains with $T_{1S}$ is increased when a sample approaches an AFM order and/or a structural transition. 
It should be noted that the $T_1$ of OVD was almost uniquely determined from a single component over the entire $T$ range, although the fraction of oxygen deficiencies is larger for OVD than for UD, revealing that the inhomogeneity of electronic states may be significant in UD close to an AFM phase and/or a structural transition. 

%--------------------  Fig.6 As-NMR Recovery curve  ---------------------
%----------------------------------------------------------------------

\begin{figure}[htbp]
\begin{center}
\includegraphics[width=0.9\linewidth]{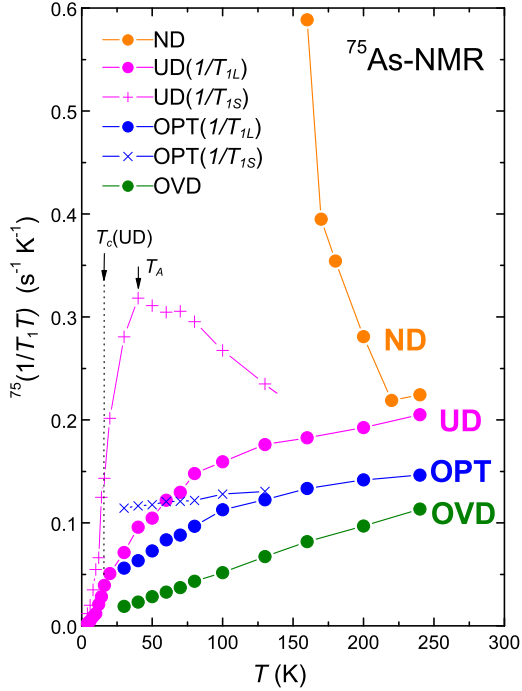}
\end{center}
\caption[]{(color online)
$T$ dependences of $^{75}$As-NMR-$(1/T_{1}T)$ for LaFeAsO$_{1-y}$ at $H\sim$ 11.97 T parallel to the $ab$-plane ($\theta=90^\circ$). $^{75}$As-$(1/T_{1L}T)$ gradually decreases upon cooling to $T_{c}$ in all samples, resembling the results of $^{57}$Fe-$(1/T_{1}T)$. 
On the other hand, $^{75}$As-$(1/T_{1S}T)$ for UD and OPT increase and stay a nearly constant upon cooling, respectively; neither of which are scaled to $^{57}$Fe-$(1/T_{1}T)$. 
}
\label{fig:AsT1}
\end{figure}

\begin{figure}[htbp]
\begin{center}
\includegraphics[width=0.9\linewidth]{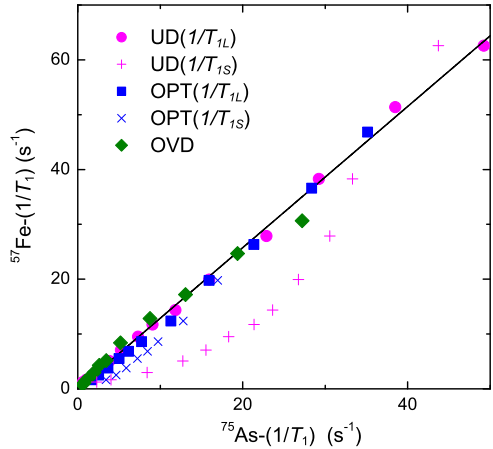}
\end{center}
\caption[]{(color online)
Plot of $^{57}$Fe-$(1/T_{1})$ versus $^{75}$As-$(1/T_{1})$ with the implicit parameter of $T$ between 30 and 240 K for UD, OPT, and OVD. The $T$ dependence of $^{57}$Fe-$(1/T_{1})$ is well scaled to that of $^{75}$As-$(1/T_{1})$ for OVD and $^{75}$As-$(1/T_{1L})$ for UD and OPT with a ratio of $^{57}(1/T_{1}T)/^{75}(1/T_{1}T)\simeq 1.3$. However, $^{75}$As-$(1/T_{1S})$ for UD and OPT samples are not scaled to $^{57}$Fe-$(1/T_{1})$, suggesting that an intrinsic property of SC phases can be probed through the result of $^{75}$As-$(1/T_{1L})$.
}
\label{fig:AsT1-FeT1}
\end{figure}

Figure~\ref{fig:AsT1} shows the $T$ dependences of $^{75}$As-NMR $(1/T_{1}T)$ at $H$=11.97 T along the $ab$-plane ($\theta=90^\circ$) for all samples. 
In OVD, $^{75}$As-$(1/T_{1}T)$ gradually decreases upon cooling to $T_{c}$, resembling the results of $^{57}$Fe-$(1/T_{1}T)$ shown in Fig.~\ref{fig:FeNMRT1}. 
$^{75}$As-$(1/T_{1L}T)$ for UD and OPT also decrease upon cooling to their $T_{c}$ similarly to the results of $^{57}$Fe-$(1/T_{1}T)$. 
In fact, Fig. \ref{fig:AsT1-FeT1} reveals that the $T$ dependence of $^{75}$As-$(1/T_{1}T)$ is well scaled to that of $^{57}$Fe-$(1/T_{1}T)$ with a ratio of $^{57}(1/T_{1}T)/^{75}(1/T_{1}T)\simeq 1.3$. 
This suggests that these $^{75}$As-NMR-($1/T_{1}$) components reveal a normal-state property inherent to UD, OPT, and OVD. 
On the other hand, $^{75}$As-$(1/T_{1S}T)$ for UD and OPT, which increase and stay a nearly constant upon cooling, respectively, are not scaled to $^{57}$Fe-$(1/T_{1}T)$, as indicated in Fig. \ref{fig:AsT1-FeT1}.
These results indicate that UD and OPT may contain unknown phases in addition to Fe-based oxypnictide superconductors. 
Here, we remark that the relaxation behaviors of $^{75}$As-$(1/T_{1S}T)$ for UD and OPT resemble those of LaFeAsO$_{1-x}$F$_{x}$ with $x=$0.04 and 0.07 \cite{Nakai2}, respectively.

In order to inspect whether or not the unknown phases are Fe-based oxypnictide phases, we measured $^{75}$As-NQR $T_1$ for UD at the $^{75}$As-NQR frequency inherent to LaFeAsO$_{1-y}$\cite{Mukuda}. 
Because the volume fraction of domains with $T_{1S}$ is increased when a sample approaches an AFM phase and/or a structural transition, it is likely that these domains arise from magnetic phases.
In order to confirm this, we next present the relaxation behavior for the nondoped sample (ND).

%**************  Fig.7  *************************************************
%****************************************************************************
%**************  Fig.8  *************************************************
%****************************************************************************

\subsection{Zero-field $^{75}$As-NMR in nondoped LaFeAsO}

\begin{figure}[tbp]
\begin{center}
\includegraphics[width=0.9\linewidth]{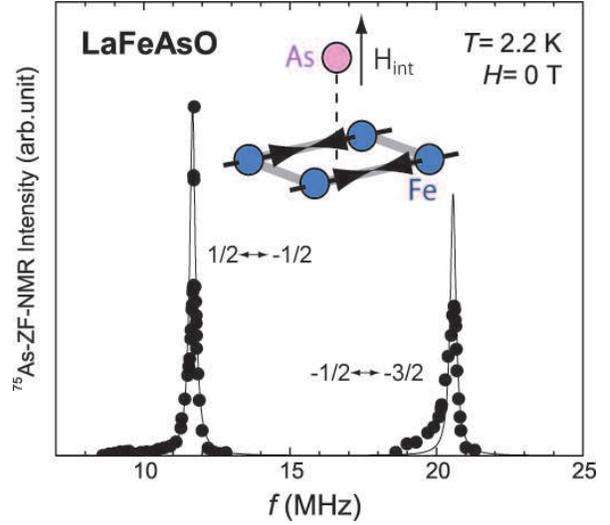}
\end{center}
\caption[]{(color online)
Zero-field $^{75}$As-NMR spectrum for nondoped LaFeAsO at 2.2 K. Two narrow spectra at 11.7 MHz (central peak) and 20.6 MHz (satellite peak) can be simulated using the NQR frequency $\nu_Q=8.8$ MHz and $H_{int}=1.60$ T at $^{75}$As along the $c$-axis, as shown by the solid curve. Note that $H_{int}$ at the As site is the off-diagonal pseudodipole field induced by the AFM moment at the Fe site, whose direction is in the $ab$-plane, which forms the stripe-type AFM structure \cite{Kitagawa}.
}
\label{fig:ZFNMRLaFeAsO}
\end{figure}

In magnetically ordered materials, the NMR spectrum can be observed at a zero field, since an internal field $H_{\rm int}$ is induced by magnetically ordered moments. Figure~\ref{fig:ZFNMRLaFeAsO} shows the $^{75}$As-NMR spectra at a zero field in the AFM state at 2.2 K for nondoped LaFeAsO (ND). The very sharp NMR spectra of ND indicate the good quality of the sample. Two narrow spectra at 11.7 and 20.6 MHz can be simulated using the NQR frequency $\nu_Q=8.8$ MHz and $H_{int}=1.60$ T along the $c$-axis, as shown by the solid curve in the figure. Note that $H_{int}$ at the As site is the off-diagonal pseudodipole field induced by the stripe-type AFM ordered moments in the $ab$-plane \cite{Kitagawa}. Interestingly, $H_{int}=1.60$ T for LaFeAsO is nearly equal to $H_{int}=1.5$ T for BaFe$_2$As$_2$ \cite{Fukazawa,Kitagawa}.  

The $^{75}$As-NMR $T_1$ of ND, which was measured in the paramagnetic state in the range $T$=150-250 K at $H=$11.97 T parallel to the $ab$-plane, was uniquely determined. $^{75}$As-NMR spectra have been published elsewhere \cite{Mukuda}. The $^{75}$As-$(1/T_1T)$ of ND is presented in Fig.~\ref{fig:AsT1}, along with those of UD, OPT, and OVD. The $^{75}$As-$(1/T_1T)$ of ND appears to be markedly enhanced toward the AFM ordering temperature $T_N\sim 145$ K, resembling the results of $^{139}$La-NMR $1/T_1$\cite{Nakai}. 
This relaxation behavior differs from those for the SC phases in UD, OPT, and OVD, which exhibit a significant decrease upon cooling. In this context, the reason for the increase in $^{75}$As-$(1/T_{1S}T)$ for UD upon cooling may be ascribed to the inclusion of some magnetic domains. We next focus on this issue. 

%-------------------- Fig.9-------------------------
%----------------------------------------------------------------------
%-------------------- Fig.10 -------------------------
\begin{figure}[tbp]
\begin{center}
\includegraphics[width=0.9\linewidth]{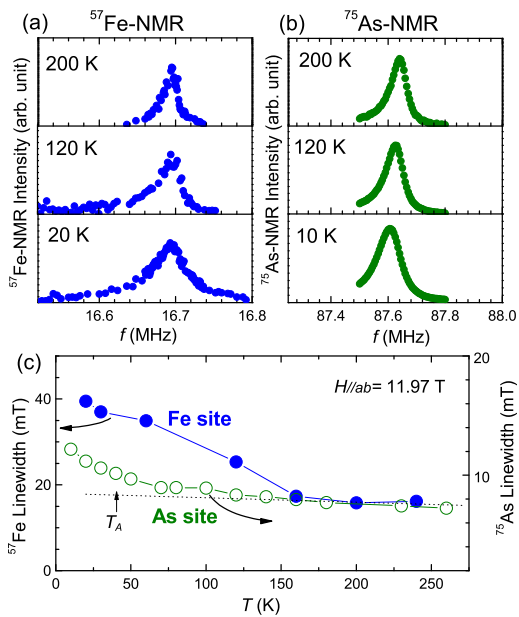}
\end{center}
\caption[]{(color online)
$T$ dependence of (a) $^{57}$Fe-NMR and (b) $^{75}$As-NMR spectra for the oriented UD sample in the field $H=$11.97 T parallel to the $ab$-plane. (c) $T$-dependence of the linewidth of these $^{57}$Fe-NMR and $^{75}$As-NMR spectra. It is noteworthy that the broadening of the NMR spectra is more significant at the Fe site than at the As site. 
}
\label{fig:linewidth}
\end{figure}
%----------------------------------------------------------------------

\section{Discussion}

\subsection{Phase separation in the underdoped sample (UD) }

The observation of two components ($T_{1S}$ and $T_{1L}$) of $T_1$ in UD indicates some local inhomogeneity in association with the density distribution of oxygen deficiencies. Namely, it is likely that magnetic domains are nucleated owing to its relatively low density of oxygen deficiencies in UD. The fraction $A_S$ of magnetic domains in UD, which exhibits the increase in $1/T_{1S}$  upon cooling, is 0.7$-$0.8, revealing that the magnetic domains appear to be separated from the SC domains when the doping level approaches an AFM phase and/or a structural transition.  Here we comment on why $^{57}$Fe-NMR-$T_1$ in UD was determined by an almost single component even though a phase separation into the magnetic and SC domains takes place. 
Figures~\ref{fig:linewidth}(a) and \ref{fig:linewidth}(b) respectively show $^{57}$Fe-NMR and $^{75}$As-NMR spectra for UD. Figure \ref{fig:linewidth}(c) shows the $T$ dependence of the linewidth of both spectra.
The $^{57}$Fe linewidth increases significantly upon cooling below 130$-$150 K, whereas the $^{75}$As linewidth increases gradually below 130$-$150 K.  These results suggest that a tiny magnetic moment is induced at the Fe site in the magnetic domains below 130$\sim$150 K. 
The further increase of $^{75}$As linewidth is seen below $T_A\sim$40 K, suggesting that the increase of magnetic-domain size upon cooling stabilizes the static short-range magnetic order (SRMO) in UD, which is corroborated by the observation of the peak in $1/T_{1S}$ at $T_A$ shown in Fig.~\ref{fig:AsT1}.  $^{57}$Fe-$(1/T_1T)$ was measured at the center of the broad $^{57}$Fe-NMR spectra that arises predominantly from the SC domains and hence was almost uniquely determined.
On the other hand, since the broadening of the $^{75}$As-NMR spectra is much less than that of the $^{57}$Fe-NMR spectra and hence the center of the $^{75}$As-NMR spectra arises from both the magnetic and SC domains, $^{75}$As-$(1/T_1T)$ includes two components, $1/T_{1S}$ and $1/T_{1L}$, corresponding to the magnetic and SC domains, respectively.   
Note that similar results were also reported in previous works on underdoped LaFeAsO$_{1-x}$F$_{x}$($x=$0.04)\cite{Nakai,Nakai2}, in which it was claimed that $T_A$ corresponds to the maximum of resistivity, implying the crossover from a bad metal to a somewhat better metal. We consider that the  tiny moments induced in the magnetic domains below $\sim$ 150 K may develop SRMO below $T_A$. Interestingly, the $1/T_{1S}$ of the magnetic domains decreases markedly below $T_c$, as shown in Fig.~\ref{fig:AsT1}, which is presumably due to an SC proximity effect in association with the mixture of  magnetic and SC domains. In this context, $^{57}$Fe-NMR studies play a vital role in deducing the intrinsic normal-state properties of SC domains in Fe-based superconductors. 

%--------------------  Fig.11   ---------------------
\begin{figure}[htbp]
\begin{center}
\includegraphics[width=0.9\linewidth]{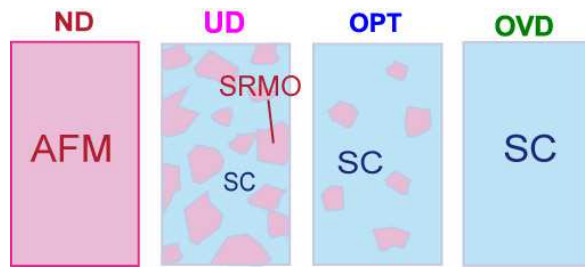}
\end{center}
\caption[]{(color online)
Illustration of the phase separation into magnetic and SC domains in UD and OPT.  As oxygen deficiencies are increased when going from UD to OVD, the fraction of magnetic domains becomes smaller in OPT and disappears in OVD, since their spatial distribution of electronic states becomes less prominent. In UD, short-range magnetic order (SRMO) in the magnetic domain is suggested from the line broadening of NMR spectra. However no such broadening of the spectra was confirmed for OPT, probably due to the very small internal field. 
}
\label{PhaseSeparation}
\end{figure}
%----------------------------------------------------------------------

\subsection{Phase diagram of LaFeAsO$_{1-y}$}

The phase diagram of LaFeAsO$_{1-y}$ in Fig. ~\ref{phasediagram} shows the evolution from AFM to SC as a function of the $a$-axis length, which becomes shorter as oxygen deficiencies are introduced into the LaO layer. 
As a result of the systematic $^{57}$Fe-NMR and $^{75}$As-NMR measurements, we reveal the phase transition between AFM and SC as a function of oxygen deficiencies in LaFeAsO$_{1-y}$, as illustrated in Fig.~\ref{PhaseSeparation}.
As the oxygen deficiencies are increased when going from UD to OVD, the fraction of magnetic domains becomes smaller in OPT and disappears in OVD, since their spatial distribution of electronic states becomes less prominent. 
The AFM order in most mother materials of Fe pnictides takes place slightly below the temperature of the first-order structural phase transition from tetragonal to orthorhombic \cite{Cruz,Luetkens,KotegawaP}. 
Accordingly, it is likely that the local inhomogeneity of oxygen deficiencies in UD is responsible for the phase separation into magnetic and SC domains.  
The features presented here resemble the phase diagram of LaFeAsO$_{1-x}$F$_{x}$ revealed by $\mu$SR\cite{Luetkens}. A phase separation has also been reported for the sample near the phase boundary between the AFM and SC states\cite{STakeshita}.

\subsection{Normal-state properties of electron- and hole-doped Fe pnictides}

\begin{figure}[htbp]
\begin{center}
\includegraphics[width=1\linewidth]{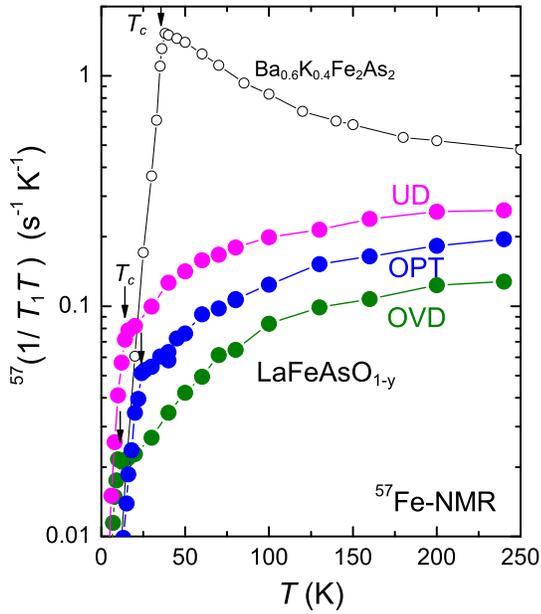}
\end{center}
\caption[]{(color online)
$T$ dependences of $^{57}$Fe-$(1/T_{1}T)$ for electron-doped LaFeAsO$_{1-y}$ and hole-doped Ba$_{0.6}$K$_{0.4}$Fe$_2$As$_2$(cited from ref. 24). $1/T_1T$ in the electron-doped LaFeAsO$_{1-y}$ decreases upon cooling, revealing that AFM spin fluctuations are not dominant, even approaching the vicinity of the AFM phase.  By contrast, the $1/T_1T$ of the optimally hole-doped Ba$_{0.6}$K$_{0.4}$Fe$_2$As$_2$ markedly increases upon cooling to $T_c$.}
\label{fig:Ba122invT1T}
\end{figure}

Finally, we compare the normal-state properties of electron-doped LaFeAsO$_{1-y}$ and hole-doped Ba$_{0.6}$K$_{0.4}$Fe$_2$As$_2$\cite{FukazawaSC,MukudaPhysC,Yashima}. Figure~\ref{fig:Ba122invT1T}(a) shows the $T$ dependences of $^{57}$Fe-$(1/T_1T)$ in the SC domains of UD, OPT, and OVD along with that for Ba$_{0.6}$K$_{0.4}$Fe$_2$As$_2$ \cite{Yashima}. 
In the case of electron-doped LaFeAsO$_{1-y}$, it is noteworthy that $(1/T_1T)$ decreases upon cooling for all superconducting samples, revealing that AFM fluctuations are not dominant, even approaching the AFM phase.
By contrast, the $1/T_1T$ of the optimally hole-doped Ba$_{0.6}$K$_{0.4}$Fe$_2$As$_2$ markedly increases upon cooling to $T_c$.
A recent theoretical work\cite{Ikeda} appears to qualitatively explain this on the basis of a fluctuation-exchange approximation (FLEX) using an effective five-band Hubbard model; in electron-doped systems, $1/T_1T$ and spin susceptibility decrease significantly upon cooling, suggesting a band structure effect, that is, the existence of a high density of states slightly below the Fermi level. 
This is consistent with the experimental finding for electron-doped LaFeAsO$_{1-y}$ that the Knight shift in OPT of LaFeAsO$_{1-y}$ decreases upon cooling as well as $1/T_1T$ \cite{Terasaki}.
On the other hand, the $T$ dependence of $1/T_1T$ is markedly enhanced down to $T_c$ in hole-doped Ba$_{0.6}$K$_{0.4}$Fe$_2$As$_2$, whereas the $T$ dependence of the Knight shift remains constant in this $T$ range \cite{Yashima}. These results  do not appear to be understandable in terms of only a simple band-structure effect in this compound; the evolution of AFM spin fluctuations should additionally be taken into account\cite{FukazawaSC,Yashima,Ikeda}. 

On the basis of these results, we claim that $1/T_1T$ is not always enhanced by AFM fluctuations close to an AFM phase in the underdoped SC sample of Fe pnictides. 
According to the theoretical study by Kuroki {\it et al.}\cite{Kuroki}, such AFM spin fluctuations may be attributed to the multiple spin-fluctuation modes arising from nesting across the disconnected Fermi surfaces. 
In this scenario, we deduce that a better nesting condition among the Fermi surfaces is realized in Ba$_{0.6}$K$_{0.4}$Fe$_2$As$_2$ than in LaFeAsO$_{1-y}$. 
Consequently, we remark that the crucial difference between the normal-state properties of LaFeAsO$_{1-y}$ and Ba$_{0.6}$K$_{0.4}$Fe$_2$As$_2$ may originate from the fact that the relevant Fermi surface topologies are differently modified depending on whether electrons or holes are doped into the FeAs layers. 

%**************  Fig.12  ********************************************
%****************************************************************************

\begin{figure}[htbp]
\begin{center}
\includegraphics[width=1\linewidth]{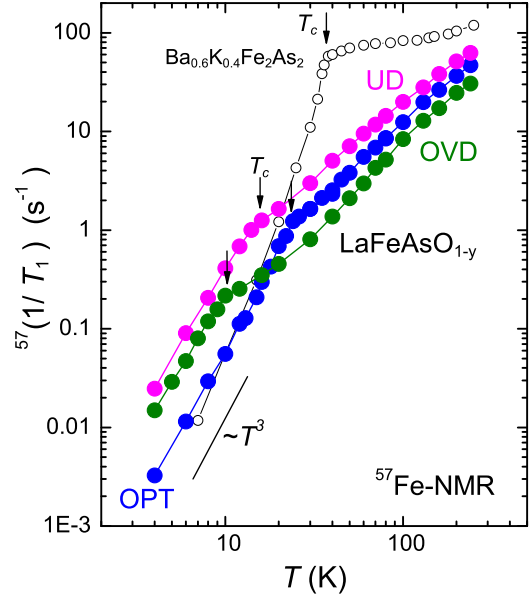}
\end{center}
\caption[]{(color online)
$T$ dependences of $^{57}$Fe-$(1/T_{1})$ for electron-doped LaFeAsO$_{1-y}$ and hole-doped Ba$_{0.6}$K$_{0.4}$Fe$_2$As$_2$(cited from ref. 24). $1/T_1$ exhibits almost $T^3$-like dependence below $T_c(H)$ regardless of the doping level in LaFeAsO$_{1-y}$, which is significantly different from the $T^5$-like dependence of Ba$_{0.6}$K$_{0.4}$Fe$_2$As$_2$\cite{Yashima}. This suggests an intimate relationship between the superconducting properties and the normal-state properties.
}
\label{fig:invT1_SC}
\end{figure}

In the superconducting state, as shown in Fig. \ref{fig:invT1_SC}, $1/T_1$ exhibits almost $T^{3}$-like dependence below $T_c(H)$ regardless of the doping level of LaFeAsO$_{1-y}$ in this $T$ range, which was significantly different from the $T^5$-like dependence of Ba$_{0.6}$K$_{0.4}$Fe$_2$As$_2$\cite{Yashima}. As reported in a recent paper\cite{Yashima}, the different behavior can be ascribed to the stronger coupling SC in Ba$_{0.6}$K$_{0.4}$Fe$_2$As$_2$ ($T_c^{max}=$ 38 K) than in LaFeAsO$_{1-y}$ ($T_c^{max}=$ 28 K). 
Therefore, it is reasonable that the difference in the superconducting properties between LaFeAsO$_{1-y}$ and Ba$_{0.6}$K$_{0.4}$Fe$_2$As$_2$ can be understood by the difference in their normal-state properties. 
In this context, we propose that the Fermi surface topology, namely a nesting condition among the disconnected Fermi surfaces, may be responsible for SC characteristics inherent to each Fe-based superconductor.
 
%**************  Fig.13  ********************************************
%****************************************************************************

%%%%%%%%%%%%%%%%%%%% Summary %%%%%%%%%%%%%%%%%%%

\section{Summary}

Systematic $^{57}$Fe-NMR and $^{75}$As-NMR/NQR studies on the oxygen-deficient Fe-oxypnictide superconductor LaFeAsO$_{1-y}$ have revealed a microscopic phase separation into SC and magnetic domains in an underdoped sample with $T_c$=20 K. As oxygen-deficiencies are increased when going from an underdoped sample to the overdoped sample, the fraction of magnetic domains becomes smaller and disappears in the overdoped sample, since the spatial distribution of electronic states becomes less prominent. It was demonstrated that $1/T_1T$ in the SC domains of the underdoped sample decreases markedly upon cooling to $T_c$ for both the Fe and As sites as well as the results in the optimally doped and overdoped samples. As a result, we consider that the decrease of $1/T_1T$ upon cooling may be a common characteristic of the normal-state property of electron-doped LaFeAsO$_{1-y}$ superconductors, regardless of the electron-doping level. This contrasts with the behavior in hole-doped Ba$_{0.6}$K$_{0.4}$Fe$_2$As$_2$, which exhibits a significant increase in $1/T_1T$ upon cooling. This crucial difference in the normal-state properties between LaFeAsO$_{1-y}$ and Ba$_{0.6}$K$_{0.4}$Fe$_2$As$_2$ may originate from the fact that relevant Fermi surface topologies are differently modified depending on whether electrons or holes are doped into the FeAs layers. 
From the comparison between the superconducting and normal-state properties, we propose that the Fermi surface topology, namely a nesting condition among the disconnected Fermi surfaces, may be responsible for SC characteristics inherent to each Fe-based superconductor.
A remaining important issue is to investigate the correlation between the electronic structure and SC characteristics in various Fe-pnictide superconductors, which may lead to a general SC mechanism of Fe-based superconductors. Experiments to investigate this issue are now in progress. 

\section*{Acknowledgements}

This work was supported by a Grant-in-Aid for Specially Promoted Research (20001004) and by the Global COE Program (Core Research and Engineering of Advanced Materials-Interdisciplinary Education Center for Materials Science) from the Ministry of Education, Culture, Sports, Science and Technology (MEXT), Japan.

%%%%%%%%%%%%%%%%%%%% bibliography %%%%%%%%%%%%%%%%%%%%%%%%%%%%%%%%%%

\end{document}